\def\preprint{1}			

\ifdefined\preprint
  \documentclass[preprint,review,12pt]{elsarticle}
\fi
\ifdefined\wordcount
  \documentclass[final,3p,times,twocolumn]{elsarticle}
\fi
\ifdefined\final
  \documentclass[final,3p,times,twocolumn]{elsarticle}
\fi

\usepackage{graphicx} 
\usepackage{color}

 \usepackage{latexsym}
 \usepackage{subfig}
 \usepackage{multirow}
\usepackage{amssymb}
\usepackage{amsthm}
\usepackage{amsmath}
\usepackage{nicefrac}

\biboptions{sort&compress}

\journal{Proceedings of the Combustion Institute}

\begin{document}

\begin{frontmatter}

\title{Enhanced DDT mechanism from shock-flame interactions in thin channels}

\author[{fir,sec}]{Hongxia Yang\corref{cor1}}
\ead{yang.hongxia@foxmail.com}

\author[sec]{Matei I. Radulescu}

\address[fir]{Fire \& Explosion Protection Laboratory, Northeastern University, Shenyang, 110004, China}
\address[sec]{Department of Mechanical Engineering, University of Ottawa, Ottawa, Ontario K1N 6N5, Canada}
\cortext[cor1]{Corresponding author:}

\begin{abstract}
We show experimentally and numerically that when a weak shock interacts with a finger flame in a narrow channel, an extremely efficient mechanism for deflagration to detonation transition occurs. This is demonstrated in a 19-mm-thick channel in hydrogen-air mixtures at pressures below 0.2 atm and weak shocks of Mach numbers 1.5 to 2. The mechanism relies primarily on the straining of the flame shape into an elongated alligator flame maintained by the anchoring mechanism of Gamezo in a bifurcated lambda shock due to boundary layers. The mechanism can increase the flame surface area by more than two orders of magnitude without any turbulence on the flame time scale. The resulting alligator-shaped flame is shown to saturate near the Chapman-Jouguet condition and further slowly accelerate until its burning velocity reaches the sound speed in the shocked unburned gas. At this state, the lead shock and further adiabatic compression of the gas in the induction zone gives rise to auto-ignition and very rapid transition to detonation through merging of numerous spontaneous flames from ignition spots. The entire acceleration can occur on a time scale comparable to the laminar flame time.
\end{abstract}

\begin{keyword}

Shock-flame interaction \sep lambda-shock bifurcation \sep deflagration-to-detonation transition \sep narrow channel 

\end{keyword}

\end{frontmatter}

\ifdefined \wordcount
\clearpage
\fi

\section{Introduction}
\label{Introduction}

The problem of shock-flame interactions has attracted much interest, as it is believed to control the transition of deflagrations to detonations (DDT) \cite{thomas2001experimental, khokhlov1999interaction, jiang2016parameterization, rakotoarison2019mechanism, rakotoarison2019detonation, yang2019dynamics}. The passage of a shock over the flame provides distortion of the flame front by the Richtmyer-Meshkov instability. Subsequent interactions with reflected shocks further deform the flame, increase the burning rate and the local gas temperature to permit transition. This deflagration to detonation transition is however highly dependent on the reactivity of the fuel, turbulence intensity and boundary conditions, such as confinement, congestion, etc...\cite{oran2007origins}. The complexity of the turbulent flame prior to transition to a detonation in typical experiments \cite{thomas2001experimental, rakotoarison2019mechanism} makes it difficult to formulate simple macro-scale conceptual models for DDT, although such attempts have been made in an ad hoc basis \cite{poludnenko2011spontaneous, saif2017chapman, rakotoarison2019mechanism, rakotoarison2019detonation} with some success. For example, numerical re-construction of Thomas' experiments of shock-flame interactions in a tube have also shown that both turbulence and boundary layer effects are extremely important \cite{gamezo2001influence,gamezo2005three} and control the burning rate in a complex manner. The reflected shock takes on a bifurcated lambda structure, where a re-circulation zone anchors the turbulent flame at the wall near the reflected shock.  

Recently, the authors \cite{yang2019dynamics} have revisited this fundamental problem of shock-flame interaction by attempting to simplify the geometry studied by Thomas, Rakotoarison and their collaborators\cite{thomas2001experimental, rakotoarison2019mechanism, rakotoarison2019detonation}. Instead of a spherical flame studied by Thomas leading to a highly convoluted turbulent flame, they considered a planar flame, such that the head-on interaction of a shock and the flame retains the macro-scale one-dimensionality of the problem. They studied hydrogen-air mixtures at low pressures (10-20kPa) in a narrow smooth channel of 1.9 cm thickness, smaller than the intrinsic flame cellular structures at those conditions, in order to probe the details of the Richtmyer-Meshkov instability. They found that the initial interaction provided a modest flame area increase consistent with Richtmyer-Meshkov deformation of an inert interface. Quite surprisingly, however, they observed that the reflected shock rapidly formed a detonation wave, in spite of the weakness of the shock strength used.    

In the present communication, we wish to elucidate quantitatively why the transition to detonations in these previously reported experiments  \cite{yang2019dynamics} was possible, in spite of the low sensitivity of the fuel-air mixture at the low pressures studied, small channel dimension, weak shock used, and absence of turbulence. As it will be shown, the thinness of the channel played the key role, by permitting highly strained flames by the anchoring mechanism of Gamezo \cite{gamezo2001influence,gamezo2005three}. Indeed, it joins other observations of increased propensity for flame acceleration and DDT in capillary tubes \cite{wu2007flame,wu2012transition, valiev2013quasi, huang2019effects, yanez2016experimental}. The extremely efficient DDT process from shock-flame interactions in narrow tubes or channels is of course extremely useful in applications where rapid DDT is desirable, such as pulse-detonation engines and detonation spray-guns, but also of relevance for accidental vapor cloud explosions, where the presence of narrow passages may provide the locus of detonation ignition in otherwise non-detonable clouds.

The present paper is organized as follows. First we report the details of the experiments. The experiment is then re-constructed by direct numerical simulation, which is followed by analysis of the detonation transition process observed in order to elucidate the physical mechanism for flame acceleration and transition to a detonation.   

\section{Experiments}
\label{Experiments}
\begin{figure*}
	\centering
	\includegraphics[width=100mm]{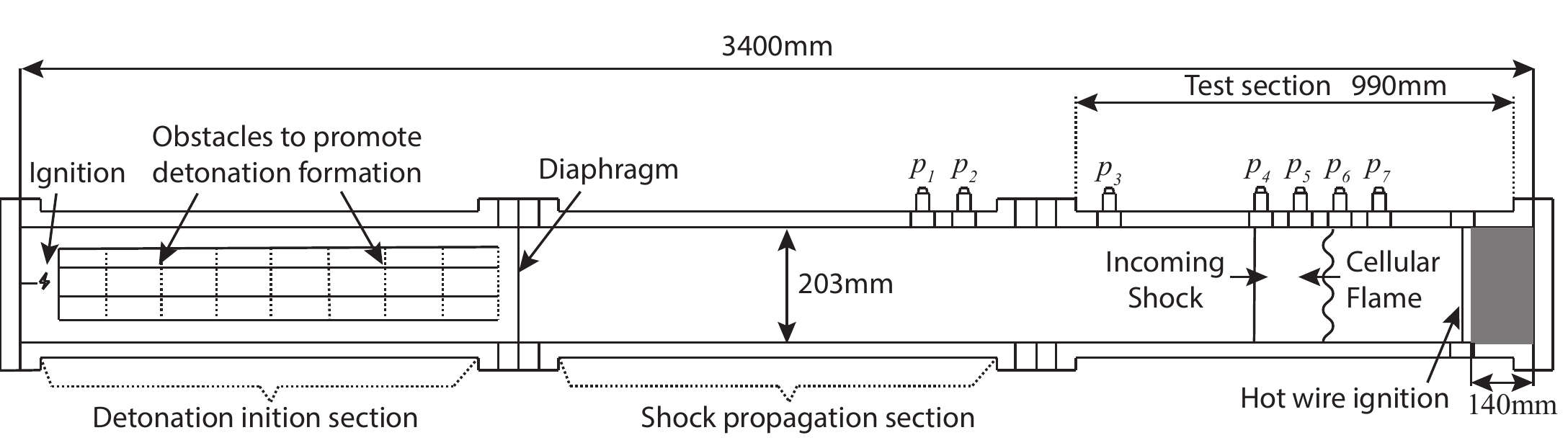}
	\caption{Schematic of the experimental setup. }
	\label{shock_tube}
\end{figure*}

The experiments were conducted in a 3400mm long, 203mm-tall and 19mm wide channel, as shown in Fig.\ \ref{shock_tube}. A reactive driver (C$_{2}$H$_{4}$ + 3O$_{2}$) gas generated a shock wave propagating to the right, while a flame was ignited at the opposite end in the test mixture, which consisted of stoichiometric hydrogen-air. The driver gas was separated from the test gas by an aluminimum diaphragm. A series of high-frequency piezoelectric PCB pressure sensors ($p_1$-$p_7$) were mounted flush on the top wall of the shock tube to collect pressure signals and the arrival of the shock. A pair of optical quality glass window was installed at the test section in order to visualize the phenomenon. The test mixture was ignited by a 0.15mm-thick tungsten wire spanning the entire channel height. When the flame has propagated a sufficient distance to acquire a quasi-stationary cellular structure, the shock wave was triggered by a detonation wave in the driver gas. The detonation then ruptured the diaphragm and transmitted a shock wave into the test gas, followed by a much weaker flame that did not participate in the experiment. The transmitted shock later traveled toward the flame and interacted head-on. A Z-type schlieren system with a field of view of 317.5mm was implemented to capture the evolution. The image sequence was recorded using a high-speed camera (Phantom v1210) with a frame rate of 59590 frames per second and an exposure time of 0.468$\mu$s. More details can be found in \cite{yang2019dynamics}.

\begin{figure}
	\centering
	\includegraphics[width=100mm]{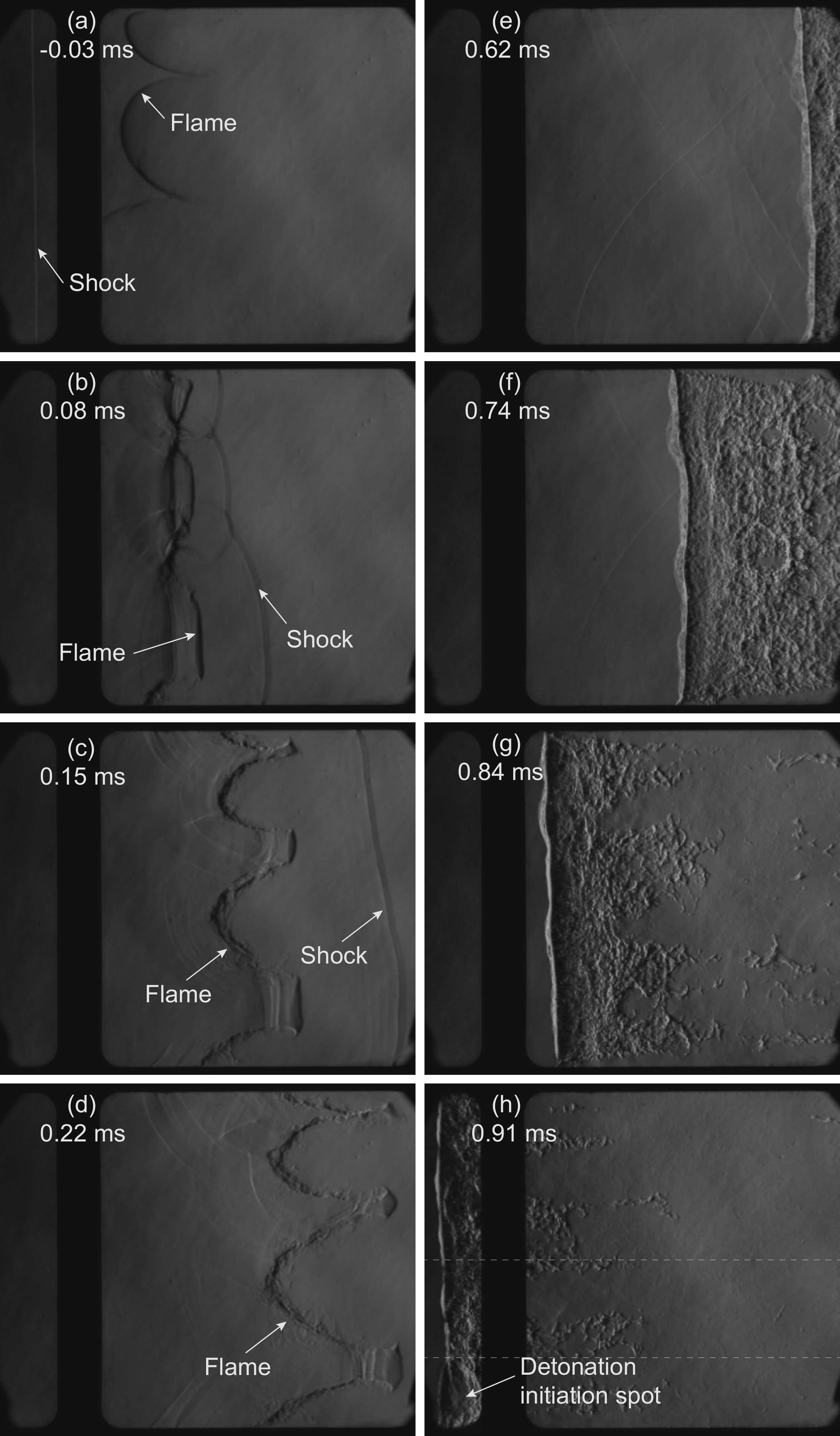}
	\caption{Schlieren image sequence of the interaction of a $M_s=1.9$ incident shock wave with a stoichiometric hydrogen-air flame at an initial pressure of 17.2kPa.}
	\label{fig:experiment_schlieren}
\end{figure}

Figure \ref{fig:experiment_schlieren} shows the detailed evolution of the interaction in a typical experiment. Here the incident shock had a Mach number of $M_{s} = 1.9$ and the test mixture was initially at 17.2kPa initial pressure. The left wall is approximately 2860 mm out of the field of view. Frame (a) shows the cellular flame structure and the incident shock before the interaction. In frame (b), the shock has passed the flame and gave rise to a transmitted shock wave. Following the passage of the incident shock, the cellular flame was reversed and pushed back to the burnt gas. Subjected to the Richtmyer-Meshkov instability, the flame acquired larger surface area indicated by longer cusps amplitudes in frame (c) and (d). In frames (e) and (f), the reflected shock traversed the flame and formed an indistinguishable shock-flame complex, similar to that observed by Thomas \textit{et al.} \cite{thomas2001experimental}. Note that the front of the shock-flame complex is relatively straight, but clearly shows that the structure has a lambda structure indicative by the finite light-gray region ahead of the main shock front \cite{mark1958interaction}. In frame (g), the shock-flame complex propagated further to the left and the flame length shrunk. A detonation ignition spot emerged near the bottom of the channel in the last frame, after which the entire front transited to detonation. The detailed video is given as supplementary material.  

\begin{figure}
	\centering
	\includegraphics[width=100mm]{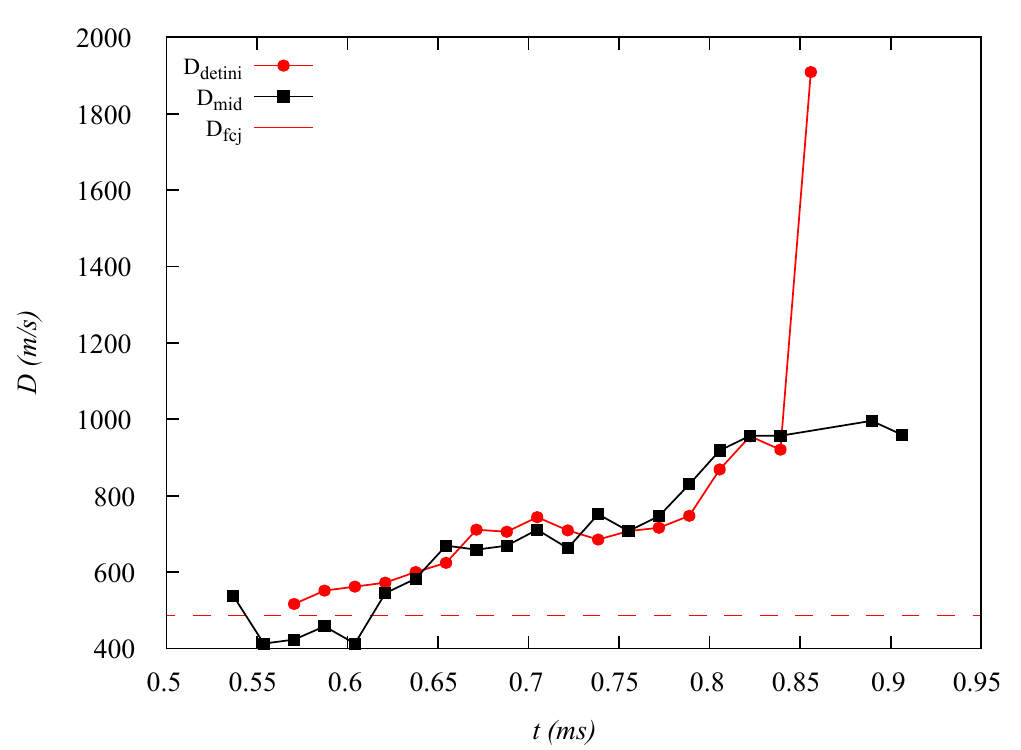}
	\caption{The speed of the reflected shock-flame complex as a function of time. $D_{fcj}$ denotes the evaluated CJ-deflagration speed after the reflected shock flame interaction. Time 0 corresponds to the first interaction of the incident shock and the cellular flame.}
	\label{fig:experiment_velocity}
\end{figure}

\begin{figure}
	\centering
	\includegraphics[width=100mm]{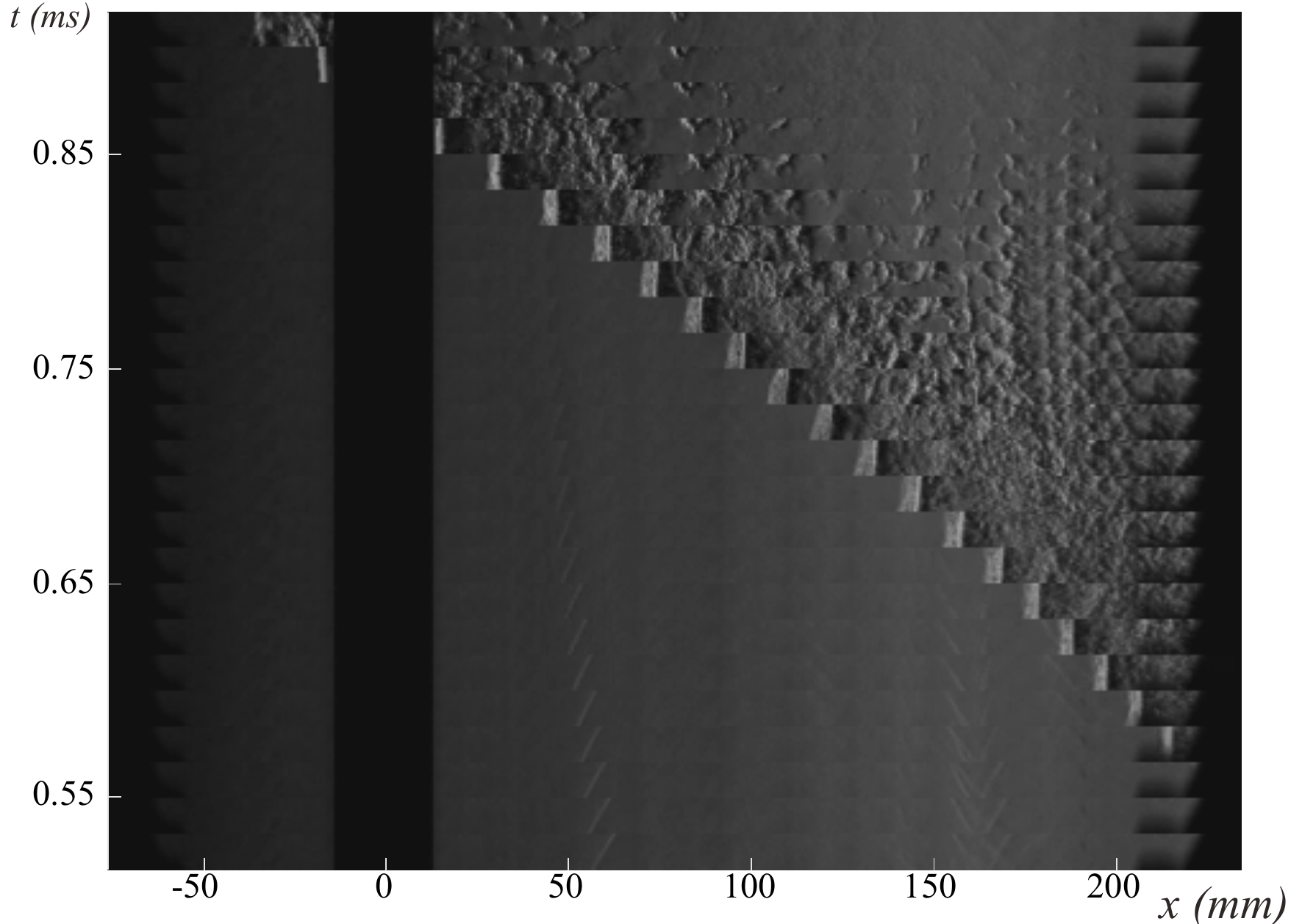}
	\caption{Space-time diagram of the shock-flame complex evolution along the bottom dashed line in frame (h) of Fig.\ \ref{fig:experiment_schlieren} with 14-pixels-wide.}
	\label{fig:experiment_xtdiagram}
\end{figure}

The speed of the reflected shock-flame complex was measured along the middle of the channel and the height where the detonation initiation spot was first emerged, as marked in the frame (h) in Fig.\ \ref{fig:experiment_schlieren}. The position of the shock flame complex leading edge was chosen to be at the main shock (thin dark band). The evolution of these two speeds is illustrated in Fig.\ \ref{fig:experiment_velocity}. The speed of the shock-flame complex at the middle of the channel and the detonation initiation spot are similar; they continuously increase until the hot spot emerged to acquire a higher velocity.      
The acceleration can be better seen on the space time diagram of Fig.\ref{fig:experiment_xtdiagram}, where the evolution of the shock-flame complex is tracked along the bottom dashed line in frame (h) of Fig. \ref{fig:experiment_schlieren} with 14-pixels-wide in each of the sequential frames after the reflected shock-flame complex were found in the schlieren image. Here, the onset of the first shock-flame interaction marks the time $t=0$ and position $x=0$.

Treating the various gasdynamic waves (shock, flames, contact surfaces and expansion waves) as gasdynamic discontinuities over which the jump conditions apply, the state behind the flame was obtained \cite{yang2019dynamics}. Near the reflecting wall the gas speed is zero, consistent with the vertical streaks in Fig.\  \ref{fig:experiment_xtdiagram} denoting the particle paths (here the trace of pockets inside the flame). The sound speed in the burned gases was found to be approximately 500m/s. By inspection of Fig.\  \ref{fig:experiment_velocity}, this signifies that the flame speed is very close to its Champman-Jouguet condition, since it propagates at the sound speed relative to the (here stationary) burned gas. It suggests that the shock-flame complex further accelerated to speeds superior to the CJ deflagration condition before transiting to a detonation.   
   
\section{Numerical Simulations}

The experiments clearly suggest that the flame evolution is three-dimensional, in spite of the thinness of the channel. While the in-plane initial Ricthmyer Meshkov flame deformation evident in Fig.\ \ref{fig:experiment_schlieren} was addressed in our earlier study \cite{yang2019dynamics}, here we focus on the flame dynamics across the thickness of the channel in the third dimension, across which the schlieren integrates. Since the Ricthmyer-Meshkov deformation visible in the schlieren photographs has a wave length much longer than the thickness of the channel, here we focus exclusively on the two-dimensional shock flame interaction in the direction of the line of sight of the experiments.

We thus formulate a two-dimensional problem in a rectangular domain, for which we model the flame propagation and gasdynamics by the reactive Navier-Stokes equations \cite{sharpe2006nonlinear}. For computational efficiency, we assume an irreversible Arrhenius rate law. Only half of the channel width was considered by assuming a symmetric boundary condition at the bottom of the domain. The domain size is set to be 750$x_0$ by 4.2$x_0$, where $x_0=\rho_{u} S_{L} c_{p}/K$ is the characteristic length scale of the flame\cite{sharpe2006nonlinear}. The  variables $\rho_{u}$, $S_L$, \textit{K}, $c_p$ are the density, the laminar flame speed, thermal conduction coefficient, and the specific heat at constant pressure of the initial unburned gas. The top and right boundaries used no-slip adiabatic wall conditions while the left boundary has imposed flow conditions corresponding to the post shock state of the desired incident shock. It is placed sufficiently far such that reflections do not reach the flame, and only serve to initially set-up a shock wave. The equations were solved using a finite-volume in-house code developed by Sam Falle at the University of Leeds \cite{maxwell2018modelling} using a second-order accurate Godunov exact Riemann solver \cite{falle1993numerical} with adaptive mesh refinement. 

The chemical kinetic parameters were calibrated in separate calculations to capture the actual kinetics of the hydrogen-air mixture in the experimental condition. The global activation energy was determined from $E_{a}=-2R \left( \partial (\ln (\rho_{u} S_{L}))/\partial (T_{ad}^{-1}) \right)_{p,\phi}$ following Egolfopoulos \cite{egolfopoulos1990chain}, by calculating the mass burning flux ($\rho_{u} S_{L}$) for the given initial pressure $p$ and equivalence ratio $\phi$, then slightly varying its value through the substitution of a small quantity of nitrogen by argon. Here, $T_{ad}$ is the adiabatic flame temperature. The pre-exponential factor is calculated by solving for the laminar flame speed satisfying the low-Mach one-dimensional steady flame structure equations using the shooting method proposed by Travnikov \cite{travnikov1997stability}. For the mixture studied, the Lewis number was set to 1.0 according to Jomaas \textit{et al.} \cite{jomaas2007transition}. 

The flame was set to start from the right side of the channel, by imposing the profiles from the steady flame calculation, and a constant initial pressure and zero velocity along the entire channel. A base grid of 1.5 grid points per flame thickness with five grid refinement levels is used. The reaction zone and diffusion zone were enforced to have the highest refinement level, giving an effective resolution of 48 points per flame thickness. When the flame tip has propagated to the same position as in the experiment, the shock reached the flame and interacted head-on.   

\section{Results and Discussion}

In order to understand the shock-flame complex structure and the mechanism of the DDT process, a series of different initial shock Mach numbers of 1.7, 1.9, 2.1, 2.3 and 2.5 were investigated. DDT was observed for all the cases other than the case of shock initial Mach number equals to 1.7. Here, we report the result of the interaction of the shock with a initial Mach number of 1.9 with a flame. Note that the incident shock went through a long distance to reach the flame, with the influence of the boundary layer effect, the shock strength to reach the flame is $M_s= 1.45$, lower than in the experiment.

\begin{figure*}
	\centering
	\includegraphics[width=107mm]{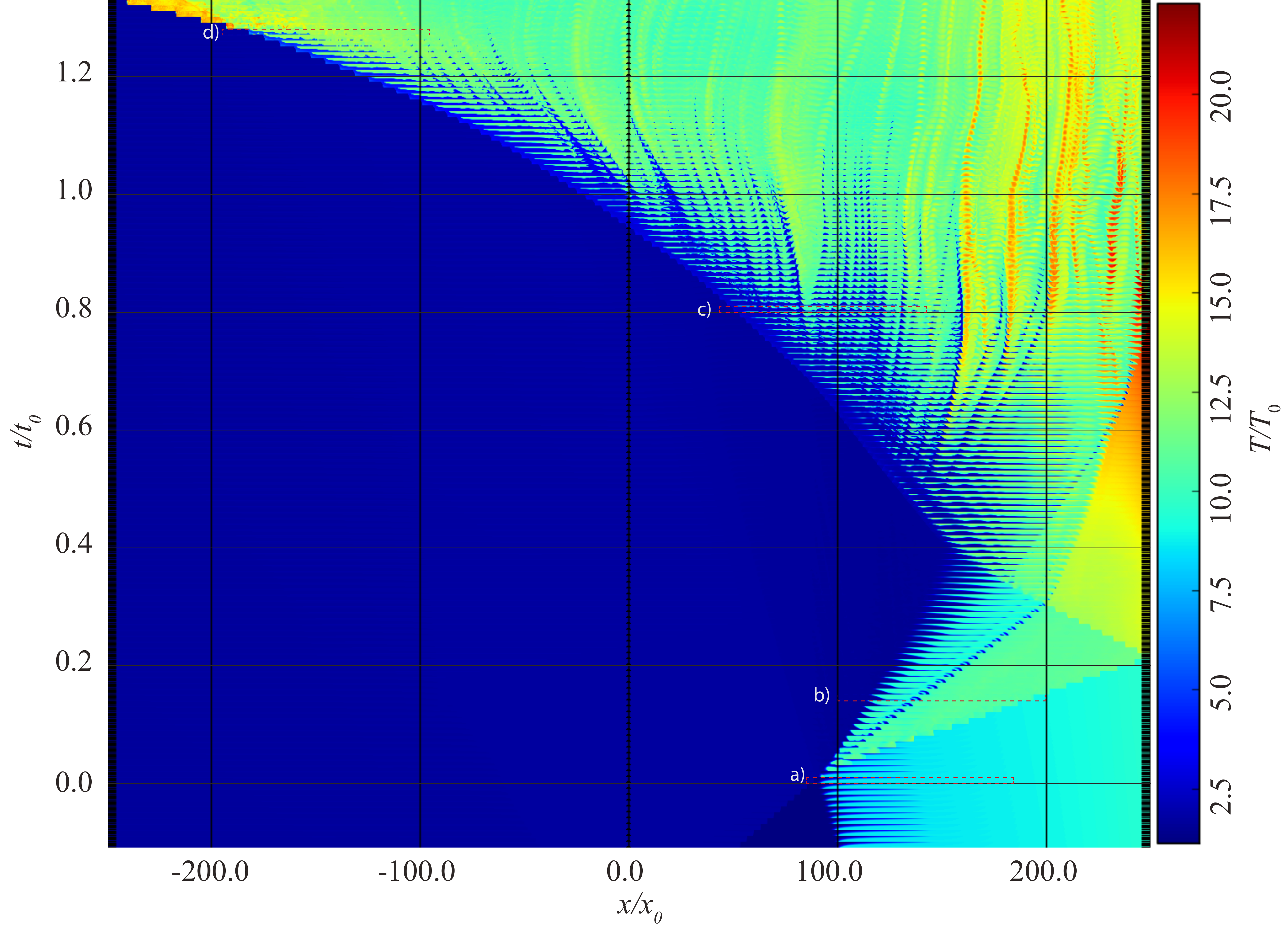}
	\caption{Space-time diagram of the evolution of the temperature profile. a)-d) are plotted in Fig. \ref{fig:sim_results}. Time 0 corresponds to the first interaction.}
	\label{fig:sim_results_xt}
\end{figure*}

\begin{figure}
	\centering
	\includegraphics[width=100mm]{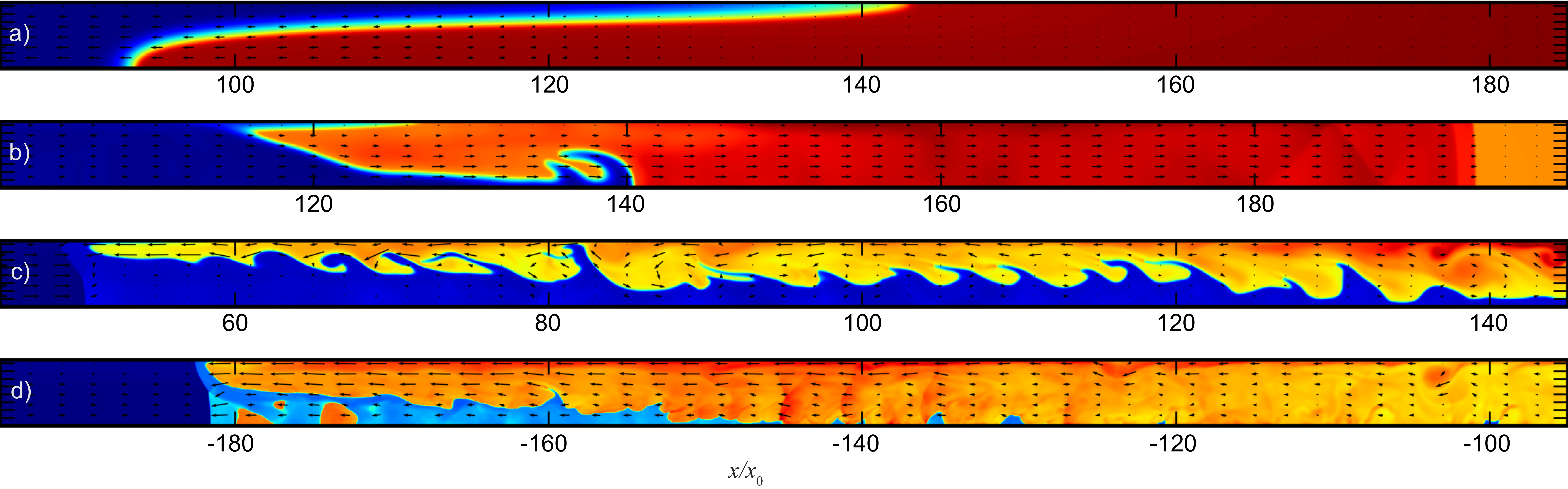}
	\caption{Temperature profiles with velocity vectors illustrating the evolution of the flame through the interactions with the incident and reflected shock.}
	\label{fig:sim_results}
\end{figure}

The flow field evolution is best understood by referring to Fig.\ \ref{fig:sim_results_xt}, together with Fig.\ \ref{fig:sim_results}, which show the evolution of the temperature profile. Fig.\ \ref{fig:sim_results_xt} shows the evolution of the temperature in the domain of interest at sequential times to form a space-time evolution of the process, while Fig.\ \ref{fig:sim_results} shows some noteworthy details. The video animation provided as supplemental material provides further clarification. Before the interaction with the leading shock, the flame has acquired an elongated finger-shape due to the no-slip condition at the wall (Fig.\ \ref{fig:sim_results}(a)). Here the surface area of the flame is 11 times larger than the channel half-width. After the shock passes over the flame, the flame gets inverted by the usual Richtmyer-Meshkov instability (Fig.\ \ref{fig:sim_results}(b)). However, when the shock reflects back to pass over the flame, the flame acquires a very elongated structure, characteristic of shocks interacting with flames from the burned towards unburned gas \cite{lafleche2018dynamicsP}. This is also accompanied by Kelvin-Helmholtz instability due to the shear illustrated in Fig.\ \ref{fig:sim_results}(c), since the lighter burned gases are accelerated forward more easily. The whirls observed in the calculations along the shear layer thus explain the long texturized flame observed in the experiments along the line of sight (Fig.\ \ref{fig:experiment_schlieren}(f)).  

The reflected shock also gives rise to a $\lambda$ structure as it interacts with the boundary layer flow of the initial incident shock \cite{mark1958interaction}. The re-circulation bubble of this shock bifurcation anchors the flame tip, as observed also in calculations by Gamezo \cite{gamezo2001influence,gamezo2005three}. It can also be seen that flow close to the boundary layer in the burnt region has acquired a relatively large velocity compared to the flow speed behind the shock in the unburned gas, which explains why the flame is anchored to the shock. The combination of flame anchoring and narrowness of the channel (which controls the initial finger shape) makes for a very efficient mechanism to strain the flame. For example, at the time of Fig.\ \ref{fig:sim_results}(c), the alligator shaped flame extends all the way to the back wall, i.e., 200 flame thicknesses $x_0$ along the channel length. The tremendous increase in flame surface area, augmented somewhat by the Kelvin-Helmholtz shear layer instability, translates in the very rapid amplification of the leading shock, clearly discernible in Fig.\ \ref{fig:sim_results_xt}. This occurs on a time scale of a single flame burnout time $t_0$. Here, $t_0 = x_0/S_L$ is the characteristic time scale of the laminar flame. This is eventually punctuated by auto-ignition events in the unburned gas internal to the flame, at contact surfaces originating from the interaction of forward facing shocks with the lead shock. This is perfectly compatible with the experimental observations, which shows that the DDT occurs within the flame brush. Also, it explains why the flame and the shock overlap, and why the flame length shrinks as the shock-flame accelerates.  

\begin{figure}
	\centering
	\includegraphics[width=100mm]{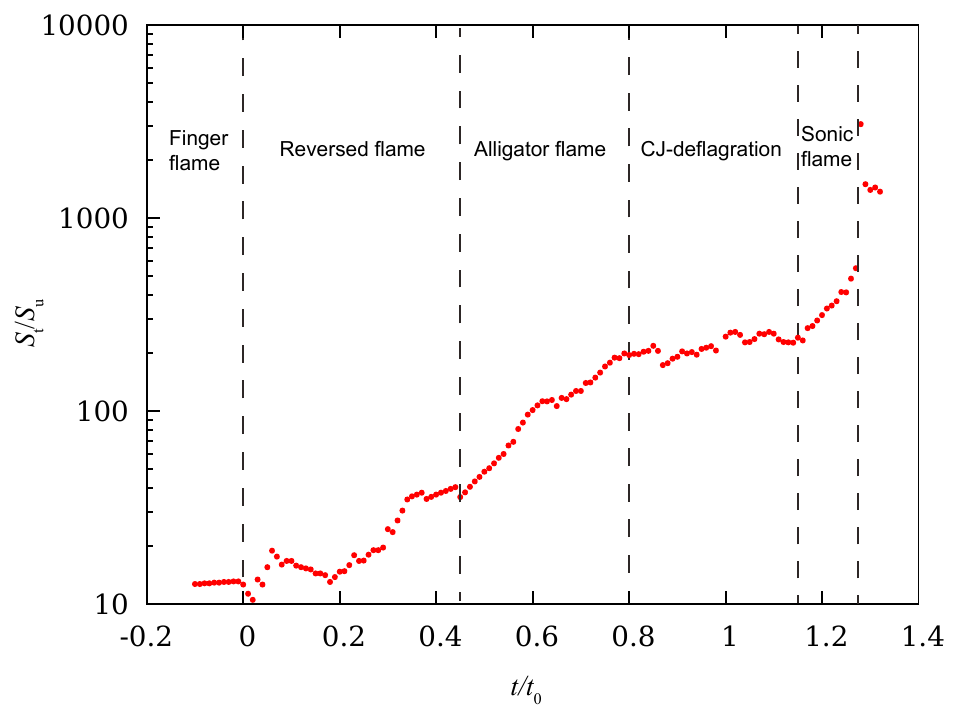}
	\caption{Effective burning velocity evolution.}
	\label{fig:burning velocity}
\end{figure}

The explosive increase in consumption rate during the flame straining is showed in Fig.\ \ref{fig:burning velocity}. It is reported as an effective burning velocity, defined as $S_{t} = \left(Y \bar{\rho}_{u}\right) ^{-1}\iint_{s} \dot \omega dxdy$, where $\bar{\rho}_{u}$ is the average density of the non-burned gas in the section of the tube occupied by the flame and $\dot \omega$ is rate of reactant mass consumption per unit volume per unit time. Time zero marks the interaction of the finger flame with the shock. Before this interaction, the effective burning velocity is approximately 12 times the laminar value, consistent with the area increase of the finger flame described above. After the interaction with the incident shock, the burning velocity increases exponentially during the flame reversal period. The exponential dependence suggests that this is not a simple linear straining of the flame by hydrodynamics alone, as expected from inert Richtmyer-Meshkov instability, but the flame consumption of non-reacted material contributes to the flame elongation, similar to the mechanism of finger flame acceleration \cite{valiev2013quasi}. 

The re-shock of the flame by the reflected shock at approximately 0.4$t_{0}$ marks a momentary reduction in the rate of acceleration, followed by a second exponential increase in consumption rate while the alligator-shaped flame further elongates. By the end of this stage, at approximately 0.8$t_{0}$, the flame has increased its consumption rate by nearly 200! This appears to be entirely due to the elongation of the flame surface area, as the consumption rate is comparable with the area increase.  The following saturated stage during which the flame consumption rate no longer increases at the same rate corresponds to a Chapman-Jouguet deflagration, as shown below. This is punctuated by a rapid run-away process at 1.15$t_{0}$.

\begin{figure}
	\centering
	\includegraphics[width=100mm]{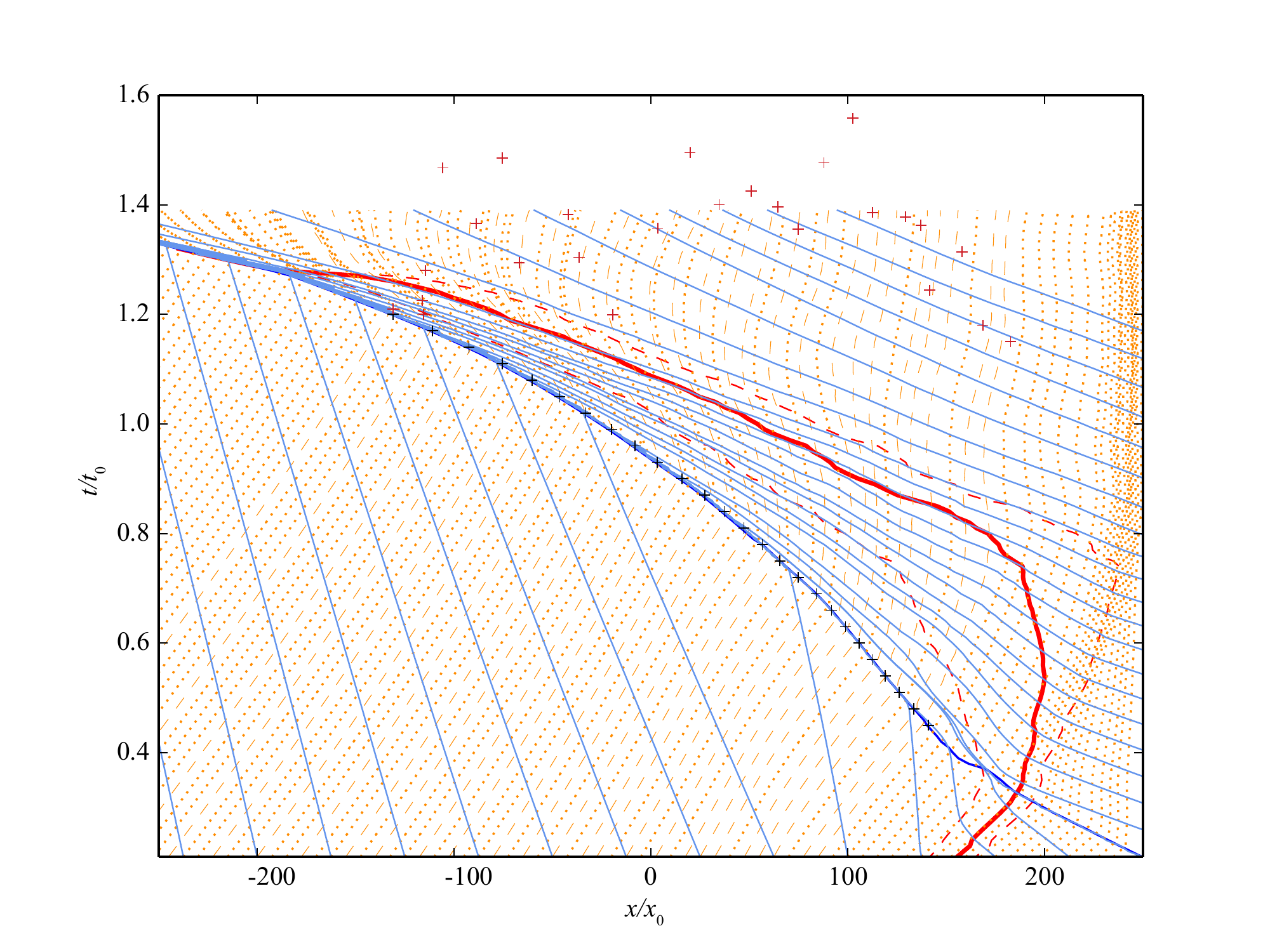}
	\caption{Characteristic space-time plot of the one dimensional Favre-averaged flow field evolution; the orange dotted and dashed lines represent the particle path; the light blue lines are the $C^{-}$ characteristics; the red line is the half reaction position of the flame, the red dashed lines indicating the $10\%$ and $90\%$ of the consumption of reactant; the black and red '+' indicates the auto-ignition time for chosen particles along the bottom of the domain.}
	\label{fig:xt_favre}
\end{figure}

At high effective burning velocities, compressible effects become important.  To clearly understand their effect on the flame acceleration, we have further analyzed our numerical simulations by re-constructing the dynamics of pressure waves, marked by the trajectory of characteristics in a one-dimensional representation of the phenomenon.  The various two-dimensional fields were Favre-averaged in the usual way across the transverse direction in order to obtain the effective one-dimensional fields \cite{radulescu2007hydrodynamic}. The evolution of one-dimensional Favre-averaged particles paths, $C^{-}$ characteristics, shock trajectory and the flame brush marked by $10\%$, $50\%$ and $90\%$ iso-contours are shown in a space-time diagram in Fig.\ \ref{fig:xt_favre}. More clearly than in Fig.\ \ref{fig:sim_results_xt}, following the interaction of the reflected shock-flame interaction, the flame's length grew until 0.8 $t_{0}$, at which it occupied the largest distance until the end wall. At this stage, the flame has become a Chapman-Jouguet deflagration: the tail of the flame, marked by the  $90\%$ iso-contour is parallel to the $C^{-}$ characteristics. 

\begin{figure}
	\centering
	\includegraphics[trim={0.2cm 0.2cm 0.2cm 0.7cm},clip,width=100mm]{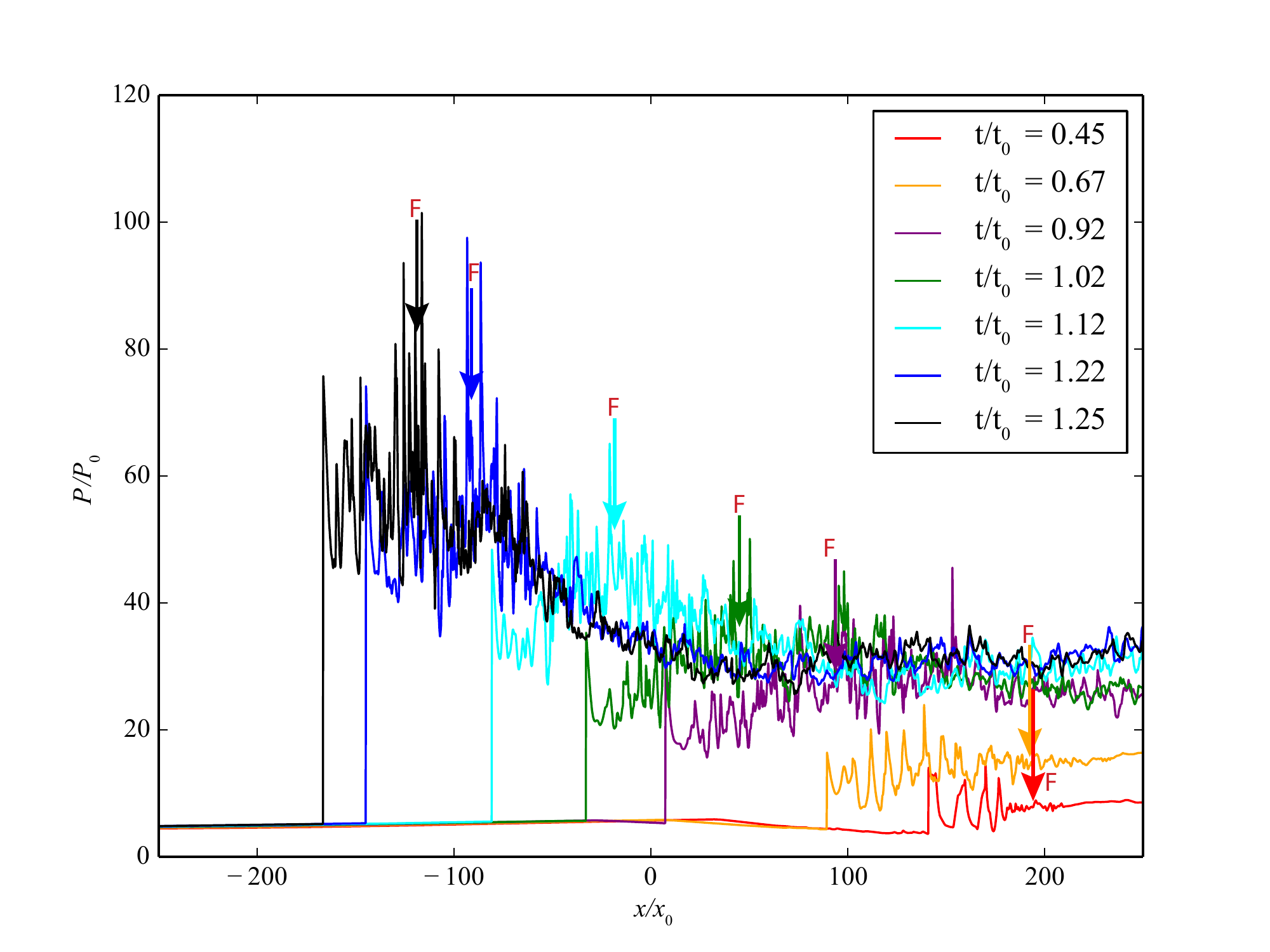}
	\caption{Pressure profiles along the bottom of the domain; arrows indicate the position of the 50\% reactant consumption.}
	\label{fig:pressure profile}
\end{figure}

At this gasdynamic stage of the flame, any subsequent increase in burning velocity translates mainly in the amplification of forward facing pressure waves. Owing to the sonic character of the flame, forward facing pressure waves are in phase with the flame, thus receive rapid amplification. These are evident by the convergence of the $C^{-}$ characteristics in Fig. \ref{fig:xt_favre}, but best seen in Fig.\  \ref{fig:pressure profile}. These pressure waves are amplified while in residence in the flame.  In the last stage of this amplification, the flame front, marked by the $10\%$ iso-contour becomes in phase with the sound waves, we call this a \textit{sonic flame}.  This marks the most rapid phase of forward facing pressure wave amplification. The net effect of this amplification stage is to strengthen the lead shock.        

Interestingly, auto-ignition plays little role during the entire flame acceleration, whose burning rate is diffusively controlled through the explosive increase in the surface area of the flame. This has been confirmed by calculating the ignition delay times in the unburned gas, by including the effect of transient compression of the gas behind the shock. A linear compression ramp for each particle was found adequate to describe this compression.  With the rate of compression given, the Lagrangian energy and species equations along each particle path become ordinary differential equations \cite{radulescu2010critical}, which were solved numerically. The resulting ignition delay times are shown in Fig.\ \ref{fig:xt_favre}. They are found to be a few times longer than the residence time of the gas in front of the flame. It is only at the end of the acceleration process that the ignition delays become shorter than the residence time.  This is in perfect agreement with the observations of hot-spot ignition in the shocked non-reacted gas at this last stage.   

\begin{figure}
	\centering
	\includegraphics[trim={0.0cm 0.0cm 0.0cm 1.5cm},clip,width=100mm]{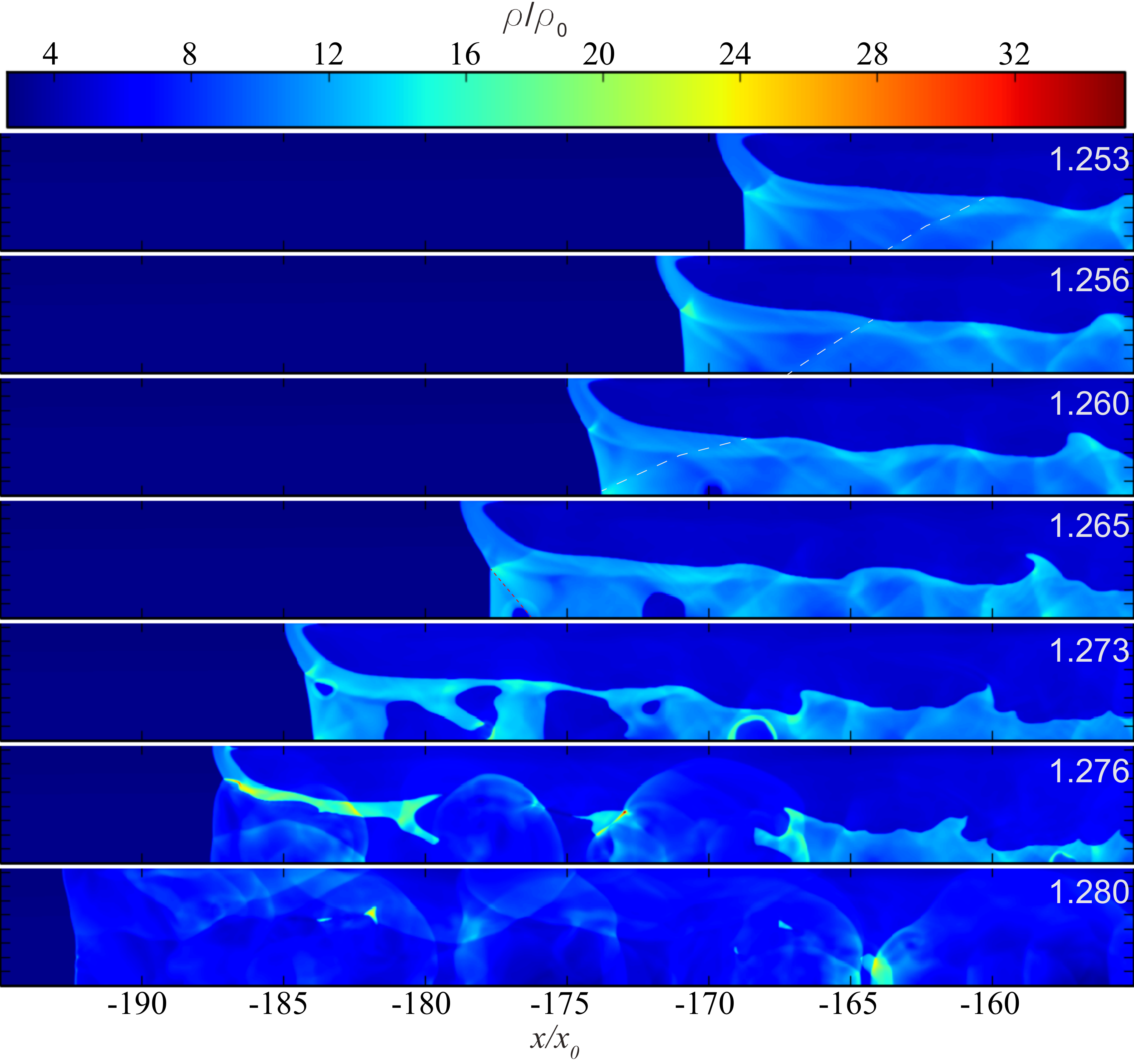}
	\caption{Density profiles illustrating the detailed formation of hot spots and the DDT process.}
	\label{fig:DDT process}
\end{figure}

The final DDT occurs very rapidly and resembles a volumetric explosion from neighboring hot spots. Fig.\ \ref{fig:DDT process} shows the details. From 1.253 $t_{0}$ to 1.255 $t_{0}$, the compression wave marked by the white dashed lines formed by the flame at an earlier time propagated to reach the precursor shock and formed a contact surface back to the right. A hot spot then emerged at the intersection of the constant surface and the bottom wall, where it has the highest temperature as the compression wave reached the bottom of the leading shock first. Then, at later times, more hot spots were formed until it finally transited to detonation.
\section{Closing remarks}
To recapitulate, the overall process of the acceleration relies uniquely on the strain of the flame. Its increase in surface area gives rise to an exponential increase in its burning velocity, similar to the acceleration of finger flames in narrow channels \cite{clanet1996tulip, valiev2013quasi}. The rapidity of the overall process, found to occur on a time scale comparable to the a single laminar flame time $t_0$ signifies that diffusive phenomena do not have sufficient time to destroy the hydrodynamically induced deformation of the flame. This permits to maintain an acceleration increasing the flame surface area by more than two orders of magnitude. The ensuing very large burning velocities bring the flame in the compressible regime, where pressure waves are phase with the elongated flame motion. This ultimately leads to the DDT. The picture of DDT is thus the classical one, with minor role played by the auto-ignition events. Rather, the formidable flame deformation and its eventual proximity to the Chapman-Jouguet limit appear as the controlling phenomena. This clarifies the emerging picture of DDT, where the proximity of the flame burning velocity to the Chapman-Jouguet value has been identified as a condition for DDT \cite{poludnenko2011spontaneous, saif2017chapman, valiev2013quasi, rakotoarison2019detonation, rakotoarison2019mechanism}.

\section*{Acknowledgments}
\label{Acknowledgments}
The authors acknowledge the financial support from the Natural Sciences and Engineering Research Council of Canada (NSERC) through the Discovery Grant ”Predictability of detonation wave dynamics in gases: experiment and model development”.  H.Y. acknowledges the finical support from the China Scholarship Council (CSC) and NSFC grant (51774068). The computations were facilitated by Compute Canada.

\bibliographystyle{elsarticle-num-PROCI}
\bibliography{references}

\end{document}